\documentclass[fleqn,usenatbib]{mnras}

\usepackage{newtxtext,newtxmath}
\usepackage{xcolor}

\usepackage[T1]{fontenc}

\DeclareRobustCommand{\VAN}[3]{#2}
\let\VANthebibliography\thebibliography
\def\thebibliography{\DeclareRobustCommand{\VAN}[3]{##3}\VANthebibliography}

\usepackage{graphicx, subfigure}	
\usepackage{amsmath}	

\usepackage{scalerel,tikz}
\usetikzlibrary{svg.path}
\definecolor{orcidlogocol}{HTML}{A6CE39}
\tikzset{orcidlogo/.pic={
 \fill[orcidlogocol] svg{M256,128c0,70.7-57.3,128-128,128C57.3,256,0,198.7,0,128C0,57.3,57.3,0,128,0C198.7,0,256,57.3,256,128z};
 \fill[white] svg{M86.3,186.2H70.9V79.1h15.4v48.4V186.2z}
 svg{M108.9,79.1h41.6c39.6,0,57,28.3,57,53.6c0,27.5-21.5,53.6-56.8,53.6h-41.8V79.1z M124.3,172.4h24.5c34.9,0,42.9-26.5,42.9-39.7c0-21.5-13.7-39.7-43.7-39.7h-23.7V172.4z}
 svg{M88.7,56.8c0,5.5-4.5,10.1-10.1,10.1c-5.6,0-10.1-4.6-10.1-10.1c0-5.6,4.5-10.1,10.1-10.1C84.2,46.7,88.7,51.3,88.7,56.8z};
}}
\newcommand\orcidicon[1]{\href{https://orcid.org/#1}{\mbox{\scalerel*{
\begin{tikzpicture}[yscale=-1,transform shape]
\pic{orcidlogo};
\end{tikzpicture}
}{|}}}}

\title[Astrometric Planets]{Detection and Characterisation of Giant Planets with \textit{Gaia} Astrometry}

\author[A.~L.~Wallace et al.]{A.~L.~ Wallace$^{\orcidicon{0000-0002-6591-5290}\,1}$\thanks{E-mail: alex.wallace1@monash.edu}, A.~R.~Casey$^{\orcidicon{0000-0003-0174-0564}\,1,2,3}$, A.~G.~A.~Brown$^{\orcidicon{0000-0002-7419-9679
}\,4}$, A.~Castro-Ginard$^{\orcidicon{0000-0002-9419-3725
}\,4}$
\\
$^{1}$School of Physics \& Astronomy, Monash University, VIC, Australia\\
$^{2}$Centre of Excellence for Astrophysics in Three Dimensions (ASTRO-3D), Melbourne, VIC, Australia\\
$^{3}$Center for Computational Astrophysics, Flatiron Institute, New York, NY 10010, USA\\
$^{4}$Leiden Observatory, Leiden University, Leiden, the Netherlands\\}

\date{Accepted XXX. Received YYY; in original form ZZZ}

\pubyear{2022}

\begin{document}
\label{firstpage}
\pagerange{\pageref{firstpage}--\pageref{lastpage}}
\maketitle

\begin{abstract}
Astrometric observations with \textit{Gaia} are expected to play a valuable role in future exoplanet surveys.  With current data from \textit{Gaia}'s third data release (DR3), we are sensitive to periods from less than 1\,year to more than 4\,years but, unlike radial velocity are not as restricted by the orbital inclination of a potential planet.  The presence and potential properties of a companion affect the primary's renormalised unit weight error (RUWE) making this a valuable quantity in the search for exoplanets.  Using this value and the fitted astrometric tracks from \textit{Gaia}, we use Bayesian inference to constrain the mass and orbital parameters of companions in known systems.  Combining this with radial velocity measurements, we show it is possible to independently measure mass and inclination and suggest HD 66141 b is a possible brown dwarf with maximum mass 23.9$^{+7.2}_{-6.4}$\,M$_{\mathrm{J}}$.  We show how this method may be applied to directly imaged planets in the future, using $\beta$-Pictoris c as an example but note that the host star is bright and active, making it difficult to draw clear conclusions.   We show how the next \textit{Gaia} data release, which will include epoch astrometry, will allow us to accurately constrain orbital parameters from astrometric data alone, revolutionising future searches for exoplanets.  Combining predicted observational limits on planet mass with theoretical distributions, we estimate the probability that a star with a given RUWE will host a detectable planet, which will be highly valuable in planning future surveys.
\end{abstract}




\section{Introduction}
The majority of exoplanets have been discovered by indirect methods such as transit and radial velocity \citep{charbonneau1999detection,mayor1995jupiter}.  Recent advances in high-contrast imaging have allowed direct detection of young giant planets \citep{marois2008direct}.  Each method comes with its own challenges and limitations: while transit is limited to short periods, direct imaging is confined to outer planets on long periods.  Radial velocity can detect planets on periods longer than the transit method but is only able to provide a lower limit on a planet's mass, and is unable to determine orbital inclination \citep{latham1989unseen}.\\
Astrometry is a technique for measuring precise stellar positions and can also be applied to searches for exoplanets.  With enough precision, it is possible to observe the change in the on-sky position of a star due to the gravitational pull of an unseen companion.  Launched in 1989, the \textit{Hipparcos} spacecraft was the first mission dedicated to precision astrometry to produce a 3-D model of the local stellar neighbourhood \citep{perryman2010}.  Combined with radial velocity measurements, its astrometric observations were also able to constrain parameters of known exoplanets discovered through radial velocity \citep{zucker2001}.  \textit{Gaia}, the successor to \textit{Hipparcos}, released its third release (DR3) in 2022 \citep{gaia2023} containing 34\,months of astrometric observations with precision of the order 0.1\,mas.  \textit{Gaia}'s prospects for the detection of exoplanets were recognised in the early stages of development \citep{bernstein1995} and later expanded upon in \citet{perryman2014astrometric}.\\
Most of the exoplanets detectable by astrometry have similar separations and periods as those detected through radial velocity.  \textit{Gaia}'s capabilities are limited for planets on wider orbits due to the long observation times required.  However, recent studies have combined \textit{Gaia} data with \textit{Hipparcos} data from over 20\,years earlier to look for discrepancies with the fitted astrometric tracks.  These differing results can indicate the presence of a long period planet at different points along its orbit.  There has been some success using this method to detect and characterise long-period planets \citep[e.g.][]{carleo24,sozzetti23} and has provided dynamical masses of planets previously detected through direct imaging \citep{snellen2018,brandt2021}.\\
Astrometry has an advantage over radial velocity in that it is not as constrained by orbital inclination.  Radial velocity is more sensitive to edge-on systems and is unable to detect face-on systems whereas astrometry is capable of detecting a system of any inclination.  Additionally, unlike radial velocity, the astrometric signal increases with the semi-major axis of a planet which means, given enough of an observation window, astrometry is capable of detecting planets over a wider range of periods.\\
This study uses data from \textit{Gaia}'s DR3 to constrain masses and orbital parameters of planet candidates detected through other methods by simulating the expected astrometric signal and applying Bayesian inference.  Since DR3 doesn't contain individual epoch observations, this study relies on the renormalised unit weight error (RUWE) which is a measure of how poorly \textit{Gaia} has fit the parallax and proper motion of the star.  This error could be indicative of an unseen and unresolved companion \citep{belokurov2020}.  Previous studies have investigated \textit{Gaia}'s sensitivity to a planet of given mass and period for a particular RUWE \citep[e.g.][]{blanco23} and found DR3 astrometry alone is not sensitive enough to detect planets around sources with RUWE<1.  This study focuses on sources with higher RUWE (>1.1) which are more likely to host detectable companions.\\
The next sections show how the RUWE is calculated and presents the details of our analysis.  Studies of systems detected by radial velocity and direct imaging are presented, placing better constraints on planet candidates.  Finally, a study is presented showing how epoch data in DR4 (expected after 2025) will allow further detection and characterization of nearby exoplanets.
\section{Astrometric Signatures}
\label{sec:gaiadr3}
\subsection{Position of Single Star}
For most stars in the \textit{Gaia} catalogue, a five parameter astrometric solution is provided consisting of right ascension ($\alpha$), declination ($\delta$), parallax ($\varpi$) and proper motion in the $\alpha$ and $\delta$ dimensions ($\mu_{\alpha}$ and $\mu_{\delta}$ respectively).  For a single star, its position at a given time can be predicted based on these parameters and the position of the \textit{Gaia} telescope.  As \textit{Gaia} scans the sky, it rotates with a period of 6 hours about an axis at an angle of 45$^{\circ}$ to the Sun.  The position of a star along the rotation direction (scanning direction) is known to a significantly higher degree of accuracy than the position perpendicular to this direction.  The position of a star can thus be said to be 1-dimensional given by:
\begin{equation}
    \tilde{x}_{AL} = d\alpha^{*}\mathrm{sin}\psi+d\delta\mathrm{cos}\psi,
    \label{eq:x_al}
\end{equation}
where $d\alpha^{*}$ is the $\alpha$ offset relative to some reference value, multiplied by cos$\delta$, $d\delta$ is the $\delta$ offset relative to the reference value and $\psi$ is the scanning angle where a value of 0 means the scanning direction is north and 90$^{\circ}$ means the scanning direction is east.  The position of a single star is given my the matrix equation:
\begin{equation}
\tilde{x}_{AL} = A\begin{bmatrix}
\Delta\alpha^{*}\\
\Delta\delta\\
\varpi\\
\mu^{*}_{\alpha}\\
\mu_{\delta}\\
\end{bmatrix},
\label{eq:pos_single}
\end{equation}
where $\Delta\alpha^{*}$ and $\Delta\delta$ are constant offsets in $\alpha^{*}$ and $\delta$ dimensions, $\varpi$ is the parallax and $\mu^{*}_{\alpha}$ and $\mu_{\delta}$ are the proper motion in $\alpha^{*}$ and $\delta$ dimensions.  The matrix $A$ is an $N\times 5$ design matrix where $N$ is the number of observation times and is given by:
\begin{equation}
    A = \begin{bmatrix}
\mathrm{sin}\tilde{\psi} & \mathrm{cos}\tilde{\psi} & \tilde{P}_{AL} & \tilde{t}\mathrm{sin}\tilde{\psi} & \tilde{t}\mathrm{cos}\tilde{\psi},
\end{bmatrix}
\label{eq:matrix}
\end{equation}
where $\tilde{P}_{AL}$ is the along-scan parallax factor which depends on the star's sky position as well as the position of the \textit{Gaia} telescope in celestial coordinates $[\tilde{x}_{G},\tilde{y}_{G},\tilde{z}_{G}]$:
\begin{equation}
\begin{split}
    \tilde{P}_{AL} = [\tilde{x}_{G}\sin\alpha-\tilde{y}_{G}\cos\alpha]\sin\tilde{\psi}+\\
    \{[\tilde{x}_{G}\cos\alpha+\tilde{y}_{G}\sin\alpha]\sin\delta-\tilde{z}_{G}\cos\delta\}\cos\tilde{\psi}
\end{split}.
\end{equation}
\subsection{Effect of a companion}
\begin{figure}
    \centering\includegraphics[width=1.0\linewidth]{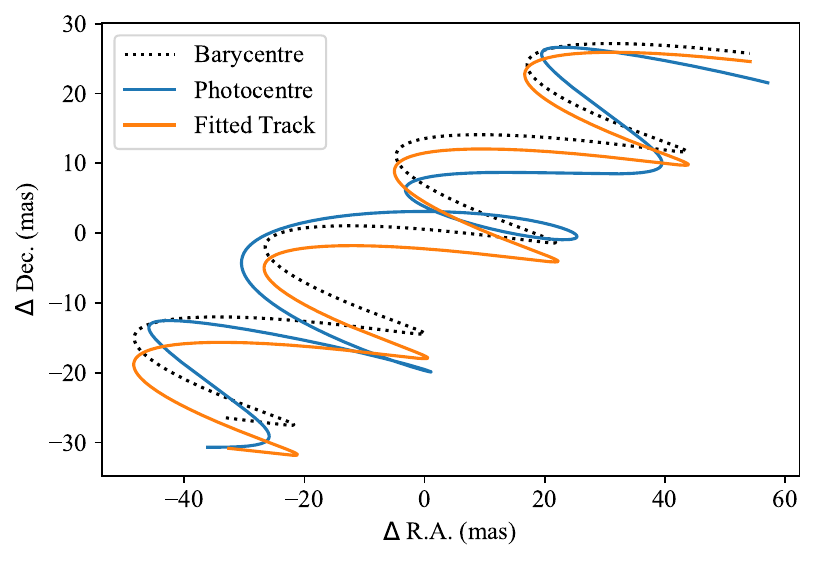}
    \caption{Example 2-D path of barycentre and photocentre of a main sequence star orbited by a brown dwarf, also showing the single star solution.}
    \label{fig:track_ex}
\end{figure}

The `position' of a star measured by \textit{Gaia} is actually the position of the centre of light (the photocentre).  For a single star, this is the same as the centre of mass (the barycentre) and is positioned according to Equation~\ref{eq:pos_single}.  For a binary system, the photocentre and barycentre are offset according to \citep{penoyre20}:
\begin{equation}
    \delta d = \frac{|q-l|}{(1+q)(1+l)}\tilde{r},
\end{equation}
where $q$ is the mass ratio, $l$ is the luminosity ratio and $\tilde{r}$ is the separation vector between the two companions.  As this study only considers planet mass companions, the companion's luminosity is negligible compared to the host star.  This equation is therefore simplified as we assume $l=0$.  The photocentre then corresponds to the position of the bright companion (the primary star).  Depending on the companion's mass and orbital parameters, this offset can significantly affect the measured position of a star when compared to a single star solution.\\
When \textit{Gaia} measures the positions, $x_{AL}$, of the photocentre along the scan, it attempts to fit a five parameter solution, assuming it is observing a single star.  This solution is given by:
\begin{equation}
    \begin{bmatrix}
\Delta\alpha^{*}\\
\Delta\delta\\
\varpi\\
\mu^{*}_{\alpha}\\
\mu_{\delta}
\end{bmatrix} = (A^{T}WA)^{-1}A^{T}W\tilde{x}_{AL}
\label{eq:obs_params}
\end{equation}
where $A$ is the $N\times 5$ matrix defined in Equation~\ref{eq:matrix} and $W$ is a weight matrix constructed by comparing the measured position to a single star track and downweighting terms which give inaccurate results.  This ensures \textit{Gaia} produces a single star track of `best fit.'  However, if the companion's mass is such that the photocentre is sufficiently offset from the barycentre, a perfect fit will be impossible.  Figure~\ref{fig:track_ex} shows a simulated example of a binary showing the motion of the barycentre and photocentre over a four year period.  The single star solution is also shown and is based on \textit{Gaia}'s observation times and scanning angles for this region of the sky.
As shown in Figure~\ref{fig:track_ex}, there are large differences between the position of the photocentre and its fitted single star track.  The single star track is also significantly offset from the path of the barycentre which means \textit{Gaia}'s measurements of the astrometric parameters will be incorrect as a result of the unseen companion.
\subsection{UWE and RUWE}
\begin{figure}
    \centering\includegraphics[width=1.0\linewidth]{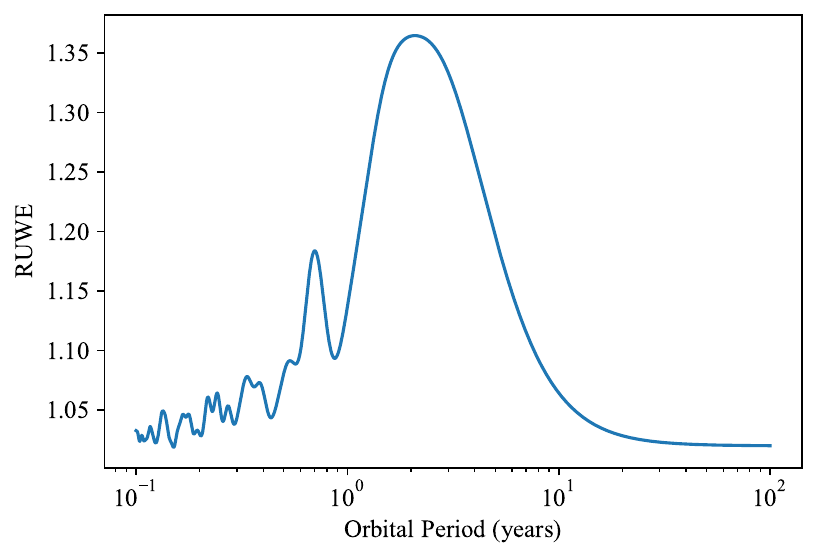}
    \caption{RUWE as a function of orbital period, for a 10\,M$_{\rm{J}}$ planet in a face-on orbit around a 1\,M$_{\odot}$ star at a distance of 50\,pc.  RUWE value depends on mass ratio and distance but the general shape of the period dependence is consistent.}
    \label{fig:ruwe_ex}
\end{figure}

Using the fitted single star track, \textit{Gaia} can measure the predicted positions of the star $x_{0}$ based on the fitted parameters, assuming a single star, using Equation~\ref{eq:pos_single}.  If the photocentre is offset due to the presence of an unseen companion, this theoretical position will differ from the observed position $x_{AL}$.  This difference is quantified by the unit weight error (UWE) of the source given by:
\begin{equation}
    UWE = \sqrt{\sum_{i}^{N} \frac{(x_{AL,i}-x_{0,i})^{2}}{\sigma^{2}(N-5)}},
    \label{eq:uwe}
\end{equation}
where $N$ is the number of observations of the source, $x_{AL,i}$ and $x_{0,i}$ are the observed and theoretical along-scan positions at the $i$-th time bin and $\sigma$ is the astrometric error of the source.  The number of degrees of freedom is given by $N-5$ because the single star astrometric solution consists of 5 parameters.\\
The UWE can be affected by factors other than a companion, namely unaccounted for modelling errors which depend on colour and brightness.  For this reason, UWE is divided by a reference value $u(G,C)$ where $G$ is the apparent G-magnitude of the source and $C$ is the difference between Bp and Rp magnitudes.  This reference value is found using the 41st percentile of UWE for sources of similar colour and brightness \citep{lindegren18}.  The division produces the `re-normalised' unit weight error (RUWE) which is available for most sources in the DR3 catalogue.  In this study, we simulate UWE from a companion's orbital parameters and compare to observed RUWE. Since in our simulations we cannot account for unmodelled errors as a function of magnitude and colour, we treat RUWE and UWE as equivalent for our study.  Effectively this means we assume high values of RUWE are exclusively due to the presence of a companion.  We use the same methods as \citet{lindegren22} to determine the dependence of UWE or RUWE on the parameters of a system.\\
A RUWE value of approximately 1 (or UWE close to the reference value) implies that any differences between $x_{0}$ and $x_{AL}$ can be attributed to astrometric errors and there is no astrometric evidence of a companion.  A high RUWE value could be indicative of an unseen companion.  In the presence of a companion, the value of RUWE depends on multiple factors such as the mass ratio, orbital period and, to a lesser extent, the viewing angles.  Mass and period have the greatest effect and the general relation between period and RUWE is shown in Figure~\ref{fig:ruwe_ex} for a 10\,M$_{\rm{J}}$ planet with a 1\,M$_{\odot}$ at a distance of 50\,pc, assuming a face-on orbit.

\noindent The relation in Figure~\ref{fig:ruwe_ex} shows a clear maximum RUWE at periods of $\sim$2.5--3\,years and a local minimum at 1\,year.  This minimum is due to the fact that, if a companion has an orbital period of 1\,year, the astrometric offset is synchronised with the star's parallax.  This has the effect of artificially raising or lowering the measured parallax but the overall track will resemble that of a single star.  The peak at 3\,years corresponds to the total length of \textit{Gaia} observations in the latest data release DR3.  Longer periods mean larger separations and greater astrometric offsets.  However, if the period is significantly longer than the timescale of observations, there is not enough motion to detect the companion.  For this reason, RUWE declines for longer periods.\\
The RUWE is also highly dependent on the mass of the companion and partly on the orbital inclination with face-on systems producing slightly higher RUWE values.  The relation between RUWE and other factors (such as eccentricity, longitude of ascending node, argument/time of periastron) depends on observation times and scanning angles.  The observation times and scanning angles are known quantities and, in the absence of detailed time-series astrometry, the RUWE is the best metric for detecting companions and constraining orbital parameters.
\section{Method to Constrain Orbital Parameters}
We constrain orbital parameters of a potential companion using the \texttt{pystan} module which allows us to build a forward model for the data and perform Bayesian inference techniques \citep{carpenter17}.  We use the observed position, parallax and proper motion from \textit{Gaia} with associated uncertainties, as well as the RUWE with an assumed uncertainty of 0.2, and simulate sets of system parameters which produce these observations, taking the effect of a companion on the measured position into account.  The system parameters constrained are: the primary mass ($M_{1}$), companion mass ($M_{2}$), period ($P$), eccentricity ($e$), inclination ($i$), longitude of the ascending node ($\Omega$), argument of periastron ($\omega$) and time of periastron ($T_p$).  The value of $T_p$ is taken to be in the range of $\pm$P/2 where a value of 0 corresponds to periastron at 2016.0 (the reference epoch.). In addition, we assume the measured position, parallax and proper motion are slightly incorrect so must also constrain the `true' values of these parameters.  Thus we have an additional five parameters: offset in R.A. and Dec. ($\Delta\alpha$ and $\Delta\delta$ respectively), true parallax ($\varpi$) and true proper motion in R.A. and Dec. ($\mu_{\alpha}$ and $\mu_{\delta}$ respectively.)\\
We calculate the posterior distribution for each of these parameters using Bayes' Theorem and assuming a uniform prior for most of the parameters.  However, for the primary mass, we use a Gaussian prior with a standard deviation of 0.5\,M$_{\odot}$ and a mean determined by the G magnitude.  This mean is calculated by assuming the measured parallax is correct, calculating the absolute magnitude and comparing to an isochrone with solar age and metallicity.  The wide uncertainty of 0.5\,M$_{\odot}$ allows for differences in age and metallicity and possible errors in distance measurement but ensures that the mass is confined to an appropriate range.  We also put hard limits of 0.08\,M$_{\odot}$ and 3\,M$_{\odot}$ to ensure the primary mass never drifts too low or too high.  Inclination has a uniform prior in cos$i$ which approximately matches measured distributions.\\
The posterior probability is calculated by simulating different sets of orbital configurations.  From the scanning law, we know the times a particular source was observed by \textit{Gaia} as well as the scanning angles, based purely on the source's location.  These observation times and angles are obtained using the \texttt{scanninglaw} python package \citep{everall21} (the nominal EDR3 version.)  For each set of orbital parameters, the position of the photocentre is calculated at each observation time and the along axis position is calculated according to Equation~\ref{eq:x_al} which allows calculation of the observed positions and RUWE according to Equations~\ref{eq:obs_params} and~\ref{eq:uwe} respectively.\\

In total, we have 13 parameters to constrain (8 orbital and 5 positional), including the primary star mass which has a Gaussian prior, but only have access to five measured positional parameters and the RUWE from \textit{Gaia}.  This makes it impossible to tightly constrain all 13 parameters but RUWE has shown itself to be mostly sensitive to mass ratio and period and, to a lesser extent, eccentricity and inclination.  In order to accurately constrain more parameters, we need more information.  The following sections detail some of the known planetary systems analysed in this work, which already have radial velocity and imaging data.  We are able to combine \textit{Gaia} observations with this information to place tighter constraints on the orbital parameters.\\
It is worth noting that RUWE can also be affected by calibration errors particularly for bright and active stars which can cause offsets in the measurement positions.  This study assumes RUWE is solely due to an unseen companion.  The masses reported for these systems are thus taken to be upper limits.
\section{Constraints on Known Systems with Radial Velocity Measurements}
\label{sec:rv}
A significant percentage ($\sim$20\%) of known exoplanets have been detected through the radial velocity method which measures the Doppler shift of the host star (from the NASA Exoplanet Archive \citep{nasa13}.)  Radial velocity is most sensitive to massive planets at short periods but can detect planets on longer periods than the transit method such that there is overlap between planets detectable by radial velocity and astrometry.  However, the radial velocity method is limited to one dimension and is only able to measure $M_{p}\mathrm{sin}i$ which places a lower limit on the planet mass.  There have been promising studies linking astrometry and radial velocity studies to constrain the system in three dimensions and obtain a accurate separate measurements of a planet's mass and inclination (e.g. \citet{tuomi09,philipot23,yahalomi23}).  For simplicity, we focus on systems with only one known planet orbiting a single star.
\subsection{Single planets orbiting single stars}
\label{sec:rv_summary}
\begin{figure}
    \centering\includegraphics[width=1.0\linewidth]{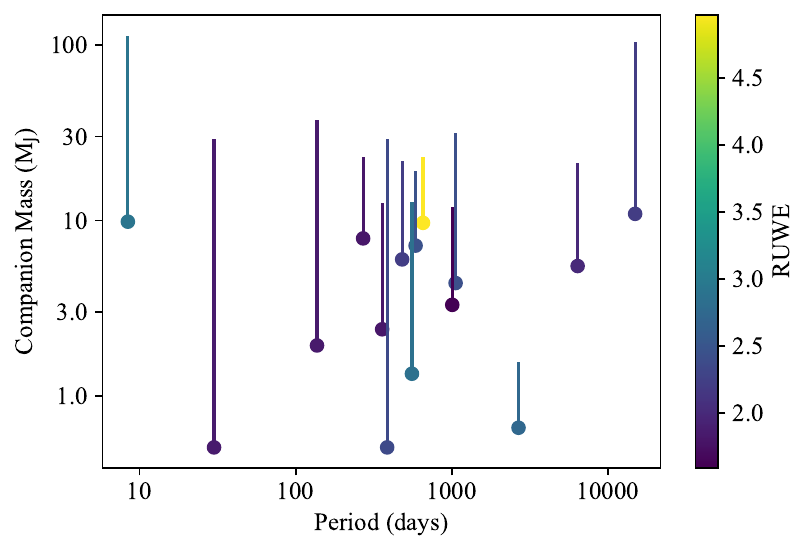}
    \caption{Minimum masses and periods calculated by radial velocity (dots) with error bars extending to most likely mass based on RUWE.  Points are coloured according to measured RUWE.}
    \label{fig:rv_summary}
\end{figure}

For this study, a sample of systems was selected from the NASA Exoplanet Archive, restricted to systems with one known star and one planet candidate discovered by radial velocity.  When considering systems of one planet orbiting a single star, there are some cases in which the planet's mass (if it were in an edge-on orbit) is too small to explain the high RUWE of the star.  In order to test this effect on many systems efficiently and identify targets of interest, a sample of radial velocity-detected planets was selected and the expected median RUWEs were calculated assuming an inclination of 90\,$^{\circ}$, using inferred mass, period and eccentricity reported from radial velocity measurements.  Focusing on systems in which the expected RUWE is significantly (>0.4) less than the measured RUWE, the mass and inclination is adjusted, keeping Msin$i$ constant until the expected RUWE agrees with the measurement.  A summary of these results is shown in Figure~\ref{fig:rv_summary}.\\
The resultant masses are summarised in Table~\ref{tab:rv_summary}.
\begin{table}
    \centering
    \begin{tabular}{c|c|c|c}
        Name & RUWE & R.V. Mass (M$_{\rm{J}}$) & RUWE Mass (M$_{\rm{J}}$)\\\hline
         4 UMa b & 1.79 & 7.9 & 23.1\\
         81 Cet b & 1.59 & 3.3 & 12.0 \\
         HD 111591 b & 2.41 & 4.4 & 31.6\\
         HD 125271 b & 2.20 & 10.9 & 104\\
         HD 141937 b & 4.97 & 9.7 & 22.9\\
         HD 162020 b & 2.90 & 9.8 & 113\\
         HD 66141 b & 2.21 & 6.0 & 21.8\\
         HD 80883 b & 1.98 & 5.5 & 21.3\\
         HIP 5763 b & 1.83 & 0.51 & 29.2\\
         $\epsilon$ Eri b & 2.72 & 0.66 & 1.56\\
         $\epsilon$ Tau b & 2.44 & 7.2 & 19.2\\
         $\gamma$ Psc b & 2.84 & 1.3 & 12.8\\
         $\mu$ Leo b & 1.79 & 2.4 & 12.6\\
         $\upsilon$ Leo b & 2.35 & 0.51 & 29.2\\
         $\xi$ Aql b & 1.83 & 2.0 & 37.1\\
    \end{tabular}
    \caption{Single-planet systems detected by R.V. with maximum companion masses suggested by RUWE, as shown in Figure~\ref{fig:rv_summary}.}
    \label{tab:rv_summary}
\end{table}
Figure~\ref{fig:rv_summary} and Tabel~\ref{tab:rv_summary} demonstrate that many of the companions orbiting stars of high RUWE have masses higher than the 13\,M$_{\mathrm{J}}$ threshold for exoplanets and could be brown dwarfs.  However, there could be other factors affecting RUWE such as calibration systematics on bright stars which cause errors in the observed positions.  In this case, the most likely masses shown in Figure~\ref{fig:rv_summary} represent upper limits.  If we were to assume the RUWE arises solely from a companion, HD 66141 b has an expected mass of $\sim$20\,M$_{\mathrm{J}}$ placing it in the mass range of brown dwarfs.  HD 125271 b and HD 162020 b had the highest possible mass ranges, both exceeding 100\,M$_{\mathrm{J}}$ while also having the shortest and longest periods respectively.  The nearby planet $\epsilon$-Eri b was has a maximum mass of 1.56\,M$_{\mathrm{J}}$, approximately 2.5 times the value from literature.  However, this system is close (3.2\,pc), bright and fast moving which could lead to calibration errors making it a difficult system to study with this method.\\

The highest RUWE in the sample (4.97) belongs to HD 141937 which has a companion with expected mass of 23\,M$_{\mathrm{J}}$ and minimum mass of 9.7\,M$_{\mathrm{J}}$.  The high RUWE of this system makes it an ideal test case for this method.  In this study, we attempt constrain the orbital parameters of a known system using the astrometric RUWE, parallax and proper motion and show how the result can be improved by adding radial velocity measurements to the model. 
\subsection{HD 141937}
\begin{figure}
    \centering\includegraphics[width=1.0\linewidth]{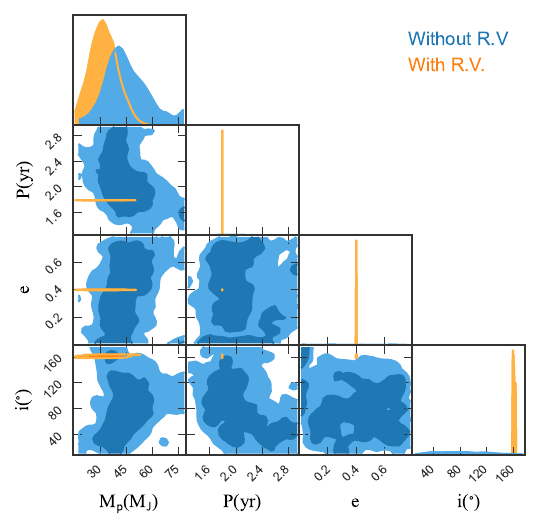}
    \caption{Posterior distributions of the mass, period, eccentricity and inclination of HD 141937 b using \textit{Gaia} data alone (blue) and adding R.V. data from \citet{udry03} (orange).  In both cases, the companion mass is indicative of a brown dwarf.  Other parameters are poorly constrained until R.V. measurements are added.}
    \label{fig:hd141937_planet}
\end{figure}

HD 141937 is a G-type main sequence star with estimated mass, age and metallicity similar to the Sun.  As stated in Section~\ref{sec:rv_summary}, this target has the highest RUWE in the sample of radial velocity-detected single planet systems, with a value of 4.97.  The companion HD 141937 b has an orbital period of 653\,days \citep{kiefer21}, well within the ideal range for astrometry with \textit{Gaia}.\\
The fitted parallax and proper motion as well as RUWE were used by the \texttt{pystan} module to obtain posteriors on the orbital parameters and true track parameters.  Initially, the radial velocity data are ignored.  The uncertainty in RUWE was assumed to be 0.2.  A second analysis was performed, this time adding the radial velocity measurements from \citet{udry03}.  The resultant posteriors for companion mass, period, eccentricity and inclination are shown in Figure~\ref{fig:hd141937_planet}.
The result in Figure~\ref{fig:hd141937_planet} shows the improvement in the result when radial velocity measurements are taken into account.  The period, in particular is poorly constrained when \textit{Gaia} measurements alone are used and the result favours periods greater than 2\,years.  When radial velocity is considered, period, eccentricity and inclination are tightly constrained.  The companion has a mass of $\sim$23.5$^{+4.5}_{-5.1}$\,M$_{\mathrm{J}}$ and the inclination is 153$\pm$2\,$^{\circ}$.  This creates minimum mass (Msin$i$) of $\sim$ 10.5$^{+2.0}_{-2.1}$\,M$_{\mathrm{J}}$ agrees with the mass calculated from radial velocity (9.7\,M$_{\mathrm{J}}$) within uncertainty.   The period is tightly constrained to 653.5$\pm$0.3\,days in contrast the poorly constrained period from \textit{Gaia} data alone.  The tightly constrained period can thus be solely attributed to radial velocity data.  The eccentricity of $\sim$0.4 also agrees with the value obtained from radial velocity.
\subsection{HD 66141}
\begin{figure}
    \centering\includegraphics[width=1.0\linewidth]{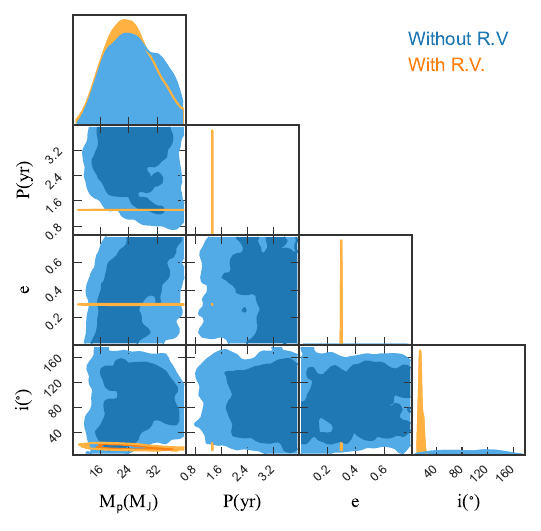}
    \caption{Posterior distributions of the  mass, period, eccentricity and inclination of HD 66141 b using \textit{Gaia} data alone (blue) and R.V. measurments from \citet{lee12} (orange).  Similar to HD 141937, the period, eccentricity and inclination are only constrained when R.V. measurements are added.}
    \label{fig:hd66141_planet}
\end{figure}

HD 66141 is an evolved K giant \citep{keenan89} at a distance of $\sim$80\,pc which has one known exoplanet with minimum mass of $\sim$6\,M$_{\mathrm{J}}$ on an almost circular orbit.  This planet was detected through radial velocity measurements between 2003 and 2011 and found to have an orbital period of $\sim$480\,days \citep{lee12}.  This star was observed by \textit{Gaia} to have a RUWE of $\sim$2.2 and, as stated in Section~\ref{sec:rv_summary}, this also suggests a companion mass in the brown dwarf range.\\
As with HD 141937, we used Stan to sample posterior probability distributions on orbital parameters first with only \textit{Gaia} data and then adding radial velocity measurements from \citet{lee12}.  A comparison of these posteriors is shown in Figure~\ref{fig:hd66141_planet}.
Figure~\ref{fig:hd66141_planet} again shows the orbital parameters are poorly constrained if \textit{Gaia} data alone is used.  The companion has a mass of 23.9$^{+7.2}_{-6.4}$\,M$_{\mathrm{J}}$ and inclination of 15$\pm$3\,$^{\circ}$.  This time, the period is constrained to 480\,days which agrees with the result from radial velocity.  The $M_{p}$sin$i$ value of 6.2\,M$_{\mathrm{J}}$ also agrees with the minimum mass presented in \citet{lee12}.  Our eccentricity of 0.30 is higher than the literature value of 0.07.  The eccentricity is difficult to constrain from \textit{Gaia} DR3 data alone and no priors were used.  This possibly caused the stan model to get stuck in a local minimum near the initial value creating a radial velocity which still closely matched the observations.  This shows a possible limitation of our method when analysing companions with low eccentricity.\\
\subsection{Summary of R.V. Results}
The results from Figures~\ref{fig:hd141937_planet} and~\ref{fig:hd66141_planet} are shown in Table~\ref{tab:rv_summary_2}
\begin{table*}
    \centering
    \begin{tabular}{c|c|c|c|c|c|c}
        & & Without R.V. & & & With R.V. & \\
        Name & Mass (M$_{\rm{J}}$) & Period (days) & Inclination ($^{\circ}$) & Mass (M$_{\rm{J}}$) & Period (days) & Inclination ($^{\circ}$)\\\hline
         HD 141937 b & 27.3$^{+7.9}_{-7.4}$ & 1140$^{+220}_{-360}$ & 103$^{+36}_{-46}$ & 23.5$^{+4.7}_{-5.1}$ & 653.5$\pm 0.3$ & 153$\pm 2$\\\\
         HD 66141 b & 25.2$^{+8.2}_{-7.3}$ & 1020$^{+310}_{-400}$ & 88$^{+47}_{-48}$ & 22.1$^{+7.9}_{-5.8}$ & 480.7$\pm 0.1$ & 17$^{+3}_{-4}$\\
    \end{tabular}
    \caption{Results from Bayesian analysis of HD 141937 and HD 66141 with and without R.V. measurements.}
    \label{tab:rv_summary_2}
\end{table*}
The classification of HD 66141 b as a brown dwarf rather than a planet would be a new result stemming from the high RUWE of the star made available in \textit{Gaia} DR3.  There could be alternative explanations for the high RUWE, such as calibration systematics, in which case this mass represents an upper limit.  There could also be an undiscovered planet.  However, another planet massive enough to effect RUWE should have already been detected through radial velocity, assuming both planets orbit in approximately the same plane.\\
Regardless of other possible explanations, these results demonstrate how the RUWE is highly dependent on the mass of a companion but, using data from \textit{Gaia} DR3 by itself, it is difficult to constrain other orbital parameters.  When combined with radial velocity, it is possible to tightly constrain period and eccentricity as well as inclination and estimate the true mass of the companion.  The result in Figure~\ref{fig:rv_summary} and the study presented in \citet{kiefer21} suggest that there are many radial velocity-detected planets which could be reclassified as brown dwarfs using this method.  However, since RUWE can also be affected by calibration errors, these mass estimates represent upper limits detailed time-series data will be required to explore this further.
\section{Planets Detected By Direct Imaging}
\label{sec:image}
The majority of planets detected by direct imaging have been found at moderate to wide separations ($\sim$10-250\,AU) \citep{currie23} beyond the capabilities of astrometry, radial velocity and transits.  However, advances in high contrast imaging and interferometry have led to the discovery of young giant planets at separations small enough to be also detected through astrometry and radial velocity \citep{nowak20}.  Studies combining direct imaging with astrometry are highly important in studies of planet formation as the mass of a directly imaged planet is typically inferred from its brightness by assuming a formation model.  If the mass can be measured independently of brightness, it will be possible to constrain planet formation models. In this section, we attempt to combine \textit{Gaia} astrometry with direct imaging observations.
\subsection{Analysis of $\beta$-Pictoris}
\begin{figure}
    \centering\includegraphics[width=1.0\linewidth]{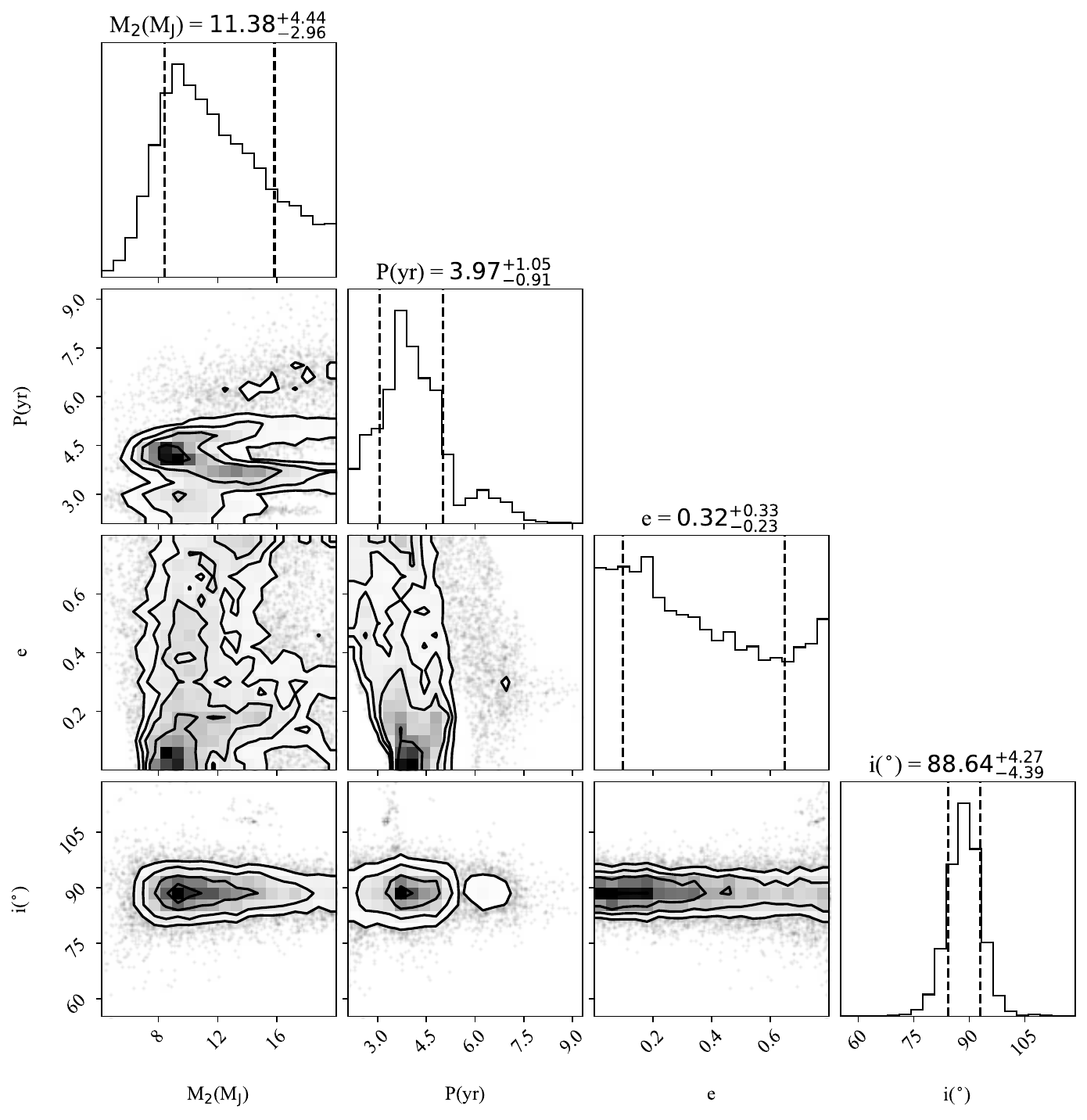}
    \caption{Posterior distributions of the $\beta$-Pic c parameters using \textit{Gaia} data and direct imaging from \citet{Lacour21}.}
    \label{fig:bPic_image}
\end{figure}

\begin{figure}
    \centering\includegraphics[width=1.0\linewidth]{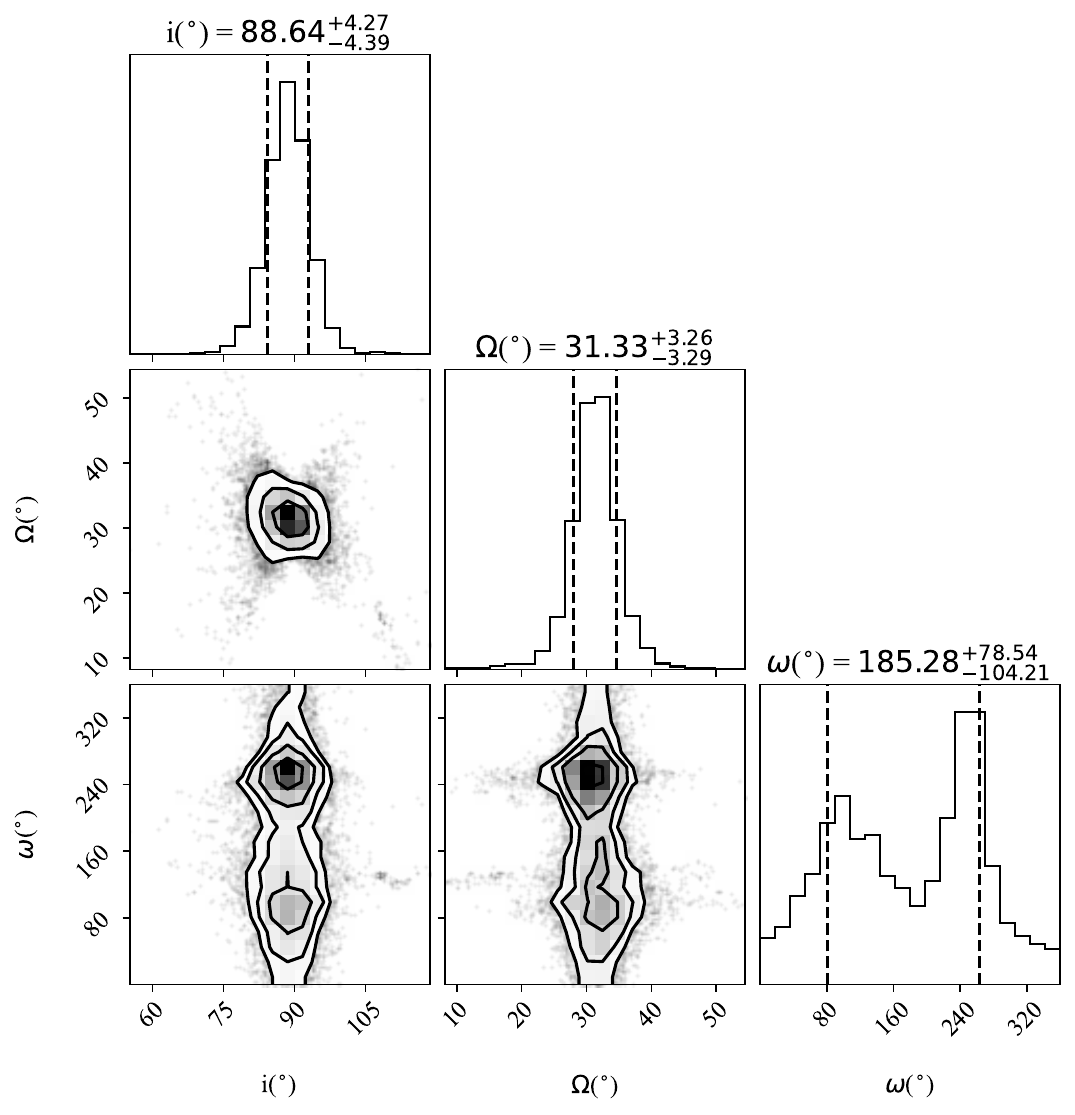}
    \caption{Posterior distributions of the inclination ($i$), longitude of the ascending node ($\Omega$) and argument of periastron ($\omega$) of $\beta$-Pic c.  $\Omega$ is tightly constrained at 31\,$^{\circ}$ and  $\omega$ shows two well-defined peaks in the distribution.}
    \label{fig:bPic_angle}
\end{figure}

$\beta$-Pictoris is a young nearby system ($\sim$20\,pc) with two planets detected by direct imaging \citep{lagrange19}.  This system is unique among direct imaging in that it is almost edge-on, with an inclination of $\sim$89\,$^{\circ}$.  However, its proximity to Earth has allowed planets on relatively short periods to be resolved by large ground-based telescopes.  The first planet detected, $\beta$-Pic b has a median period of 22.5\,years \citep{wang16} which is probably beyond the range of detectability by \textit{Gaia} astrometry.  However, a second planet $\beta$-Pic c has a period of 3.3\,years \citep{lagrange19} and has been observed through radial velocity, direct imaging and interferometry.  It is notable for having the smallest separation from its host star of any directly imaged exoplanet.  The period of $\beta$-Pic c makes this system an ideal target for studies with \textit{Gaia}.  It has a high RUWE of 3.07 which could be indicative of companions but is also a very bright target (apparent G magnitude of 3.8) and susceptible to calibration errors.  This system was chosen as a case study because it contains the only directly imaged planet ($\beta$-Pic c) with a period comparable to \textit{Gaia}'s observation window.  In this study, we attempt to constrain the mass of a planetary companion using the RUWE from \textit{Gaia} as well as direct imaging observations of $\beta$-Pic c from \citet{Lacour21}.  Including the direct imaging observations should greatly improve constraints on orbital elements.  The study from \citet{Lacour21} presented four observations of $\beta$-Pic c taken between February and December 2020 which were added to the stan model.  For simplicity, when modelling the astrometric signal, we assume only one planet as the other known planet is considered too far away to significantly affect the RUWE.  The resultant posterior distributions are shown in Figure~\ref{fig:bPic_image}.
The distributions in Figure~\ref{fig:bPic_image} show a peak in planet mass near 9\,M$_{\mathrm{J}}$ and a median of 11.38\,M$_{\mathrm{J}}$ which is slightly higher but close to the previously reported value of 8.89\,M$_{\mathrm{J}}$ \citep{Lacour21}.  The mass distribution also shows a clear tail towards higher masses.  The period has a median of 3.97\,years which is also larger than the accepted value of 3.3\,years but agrees within our uncertainty.  There is also a secondary peak in the period distribution centred around 6.5\,years which correlates with higher planet masses.  This is possible evidence of the outer planet which may have a non-negligible effect on the astrometric signal.  The star may have a larger offset due to this outer planet which, if we assume one planet, is explained by higher planet mass and longer period.  The median eccentricity closely agrees with the accepted value of $\sim$0.3 but the large uncertainty shows this parameter is difficult to constrain from limited direct imaging and astrometry alone, particularly for an edge-on system.\\
An advantage direct imaging has over radial velocity is our ability to get tighter constraints on the shape of the orbit.  Figure~\ref{fig:bPic_image} shows the inclination is tightly constrained with an uncertainty of only 4\,$^{\circ}$ which cannot be achieved through radial velocity or astrometry alone.  The images can also provide information about the longitude of the ascending node ($\Omega$) which, as it is an angle on the plane of the sky, is impossible to constrain through radial velocity.  The posterior distributions of the orbital angles (the inclination, longitude of ascending node and argument of periastron) are shown in Figure~\ref{fig:bPic_angle}.
As shown in Figure~\ref{fig:bPic_angle} the longitude of the ascending node ($\Omega$) is tightly constrained at 31\,$^{\circ}$.  As this system is known to be almost edge-on, the four observations showed the planet travelling on an almost straight line away from and towards the star.  When the inclination is taken to be 90\,$^{\circ}$, this makes calculation of $\Omega$ trivial (-1/tan$\Omega$ would be the slope of this line.)  The argument of periastron ($\omega$) shows two peaks 180\,$^{\circ}$ apart in its distribution due to some degeneracy.  The higher peak is at 260\,$^{\circ}$ and the shorter peak is at 80\,$^{\circ}$.  These two peaks arise from degeneracy between $\omega$ and time of periastron which are difficult to separate through astrometry in an edge-on system.\\

\begin{figure}
    \centering\includegraphics[width=1.0\linewidth]{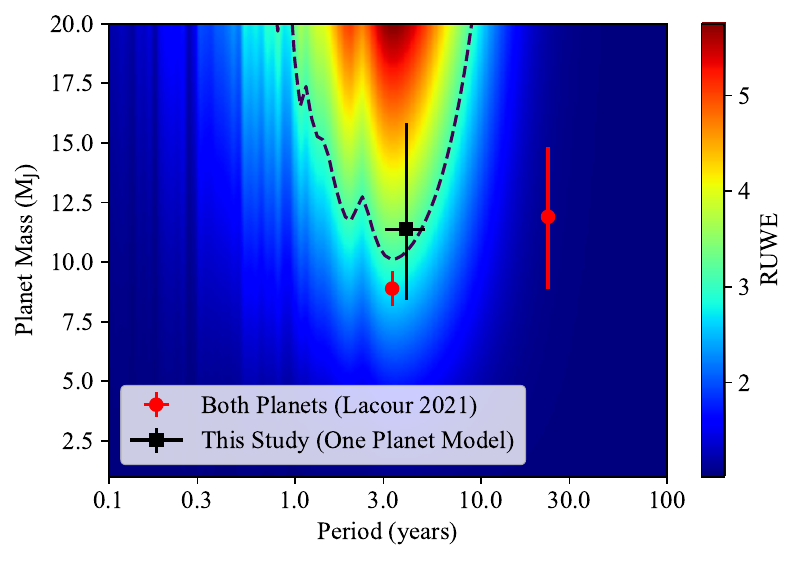}
    \caption{Expected RUWE of $\beta$-Pic as a function of planet mass and period, assuming a single companion in an edge-on orbit.  The dashed contour is the UWE reported by \textit{Gaia} and the black square is the result shown in Figure~\ref{fig:bPic_image}.  The red circles are the positions of the two planets from \citet{Lacour21}.}
    \label{fig:bPic_ruwe_contour}
\end{figure}

It is worth noting that the direct imaging data, when applied to this method, has helped constrain the period and orbital elements but the planet mass can only be constrained when RUWE is applied, assuming RUWE is mainly the result of companions.  This demonstrates the potential power of RUWE as an indicator of both the presence of a companion and that companion's mass.  It is important to investigate how much the mass of a companion affects the measured RUWE of the system.  The dependence of expected UWE on mass and period is shown in Figure~\ref{fig:bPic_ruwe_contour}, assuming a single companion in an edge-on orbit.

\begin{figure}
    \centering\includegraphics[width=1.0\linewidth]{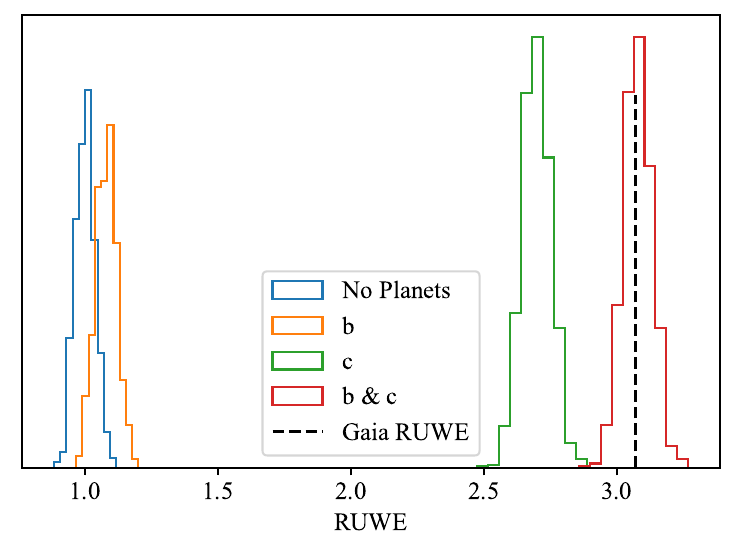}
    \caption{Distribution of UWE of $\beta$-Pic for no planets, only $\beta$-Pic b, only $\beta$-Pic c and both planets.}
    \label{fig:bPic_ruwe_dist}
\end{figure}

\noindent The simulation shown in Figure~\ref{fig:bPic_ruwe_contour} shows that the expected UWE from $\beta$-Pic c alone is $\sim$2.7 which is close to the \textit{Gaia} value of 3.07 but falls slightly short whereas the UWE from $\beta$-Pic b alone is close to 1 due to its longer period having little effect on the short term position of the photocentre.  Thus, the observed RUWE can be mostly (but not entirely) attributed to the inner planet $\beta$-Pic c.  The effect of each planet on UWE is demonstrated in Figure~\ref{fig:bPic_ruwe_dist}.  This simulates the system 1000 times to represent the possible distribution of errors so UWE is represented by a distribution for each case, rather than a single value.

\noindent The distributions shown in Figure~\ref{fig:bPic_ruwe_dist} demonstrate an increase of UWE as more planets are added and also shows the larger effect of $\beta$-Pic c.  The distribution for a single star with no planets has a peak at exactly 1 while the model with $\beta$-Pic b alone peaks at 1.1.  The model with $\beta$-Pic c alone peaks at 2.7 while the model with both planets agrees with the \textit{Gaia} reported RUWE.\\

The result from $\beta$-Pictoris demonstrates how direct imaging and astrometry can be combined for short period planets to produce better constraints on orbital parameters.  Though the brightness of $\beta$-Pic and the potential for calibration errors make our assumptions unrealistic, future interferometric surveys are expected to detect similarly short period planets around more suitable targets.
\subsection{Summary of Direct Imaging Results}
As $\beta$-Pictoris is the only system with a directly imaged planet on a $\sim$3\,year period this was our best option for combining astrometric observations from \textit{Gaia} alone with direct imaging.  Our study suggests that the RUWE can be explained by the two known planets, but the high luminosity of this system makes astrometric observations unreliable.  More success has been found for planets on longer periods using the proper motion anomaly technique, looking for discrepancies between \textit{Gaia} and \textit{Hipparcos}.  This method has predicted the existence of a second planet around 51 Eri \citep{deRosa19} and a planet around AF Lep, which was later confirmed through direct imaging \citep{derosa23} but this analysis is beyond the scope of our study.\\
$\beta$-Pic c was first detected through radial velocity, then observed through optical interferometry.  It is expected that in the next decade, advances in interferometry with at the VLTI, coupled with the construction of the E-ELT will allow direct detections of young planets on short orbital periods.  Some of the host stars may be more suitable targets of \textit{Gaia} studies than $\beta$-Pic which will make this method more viable.  However, a major step forward in \textit{Gaia} analysis will come with the next data release, DR4.
\section{Epoch Astrometry in DR4}
\label{sec:epoch}
\begin{figure}
    \centering\includegraphics[width=1.0\linewidth]{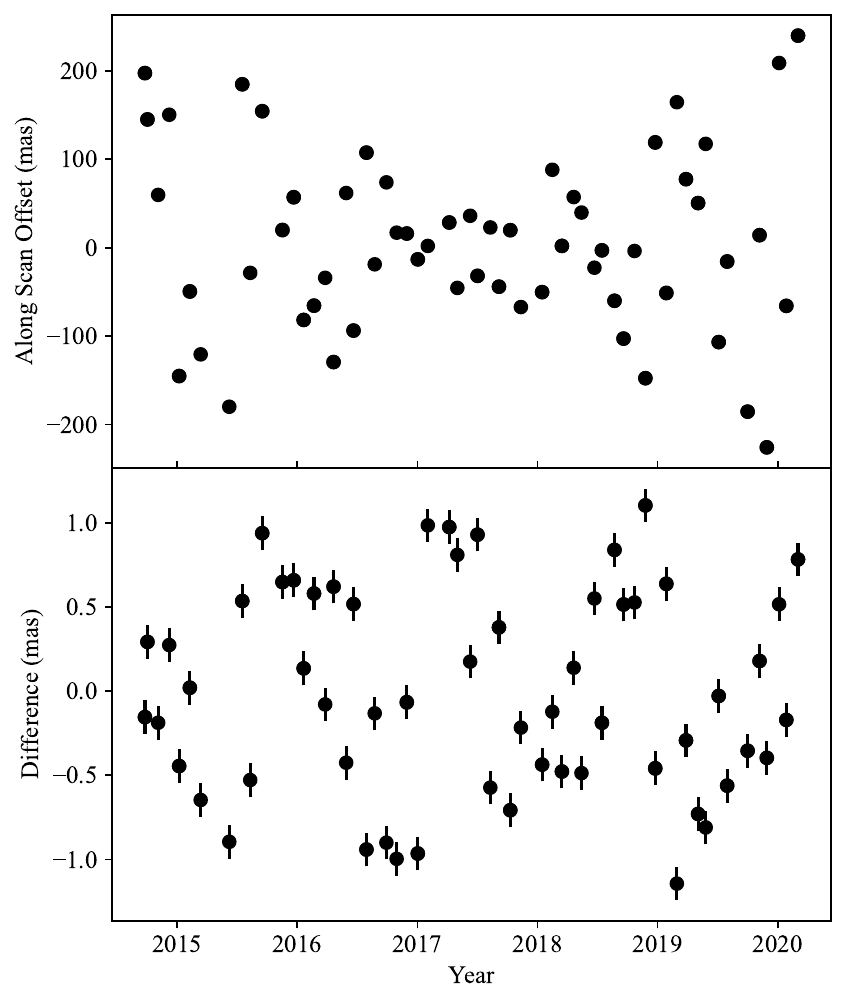}
    \caption{\textbf{Top:} The astrometric offset along the \textit{Gaia} scan direction for the photocentre of a 1\,M$_{\odot}$ star with a single 10\,M$_{\mathrm{J}}$ on a 3.5\,year orbit at 65\,$^{\circ}$ inclination, covering time range of DR4.  \textbf{Bottom:} The difference between the measured offset and the single star fit, assuming constant astrometric error of 0.1\,mas.}
    \label{fig:epoch_sim}
\end{figure}

\begin{figure}
    \centering\includegraphics[width=1.0\linewidth]{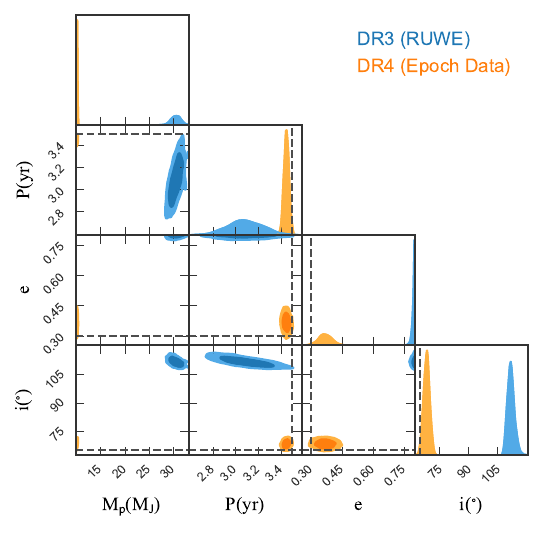}
    \caption{Planet parameters of simulated source using only fitted track parameters and RUWE from DR3 (blue) and using expected epoch data from DR4 (orange).  Black dashed lines are simulated values.}
    \label{fig:bPic_epoch}
\end{figure}

\begin{figure}
    \centering\includegraphics[width=1.0\linewidth]{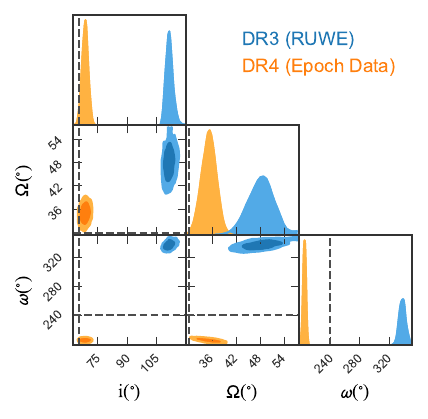}
    \caption{Orbital inclination ($i$), longitude of ascending node ($\Omega$) and argument of periastron ($\omega$) fits of simulated source without epoch data (blue) and with epoch data (orange).}
    \label{fig:bPic_epoch_ang}
\end{figure}

As stated in Section~\ref{sec:gaiadr3}, \textit{Gaia} DR3 contains fitted track parameters such as parallax and proper motion as well as the RUWE describing how poorly these parameters fit to the track.  This value has been shown to be indicative of a hidden companion but, as shown in Sections~\ref{sec:rv} and~\ref{sec:image}, extra information is required to more tightly constrain the companion's mass and orbital parameters.  The upcoming \textit{Gaia} DR4 (due to be released in mid--late 2026) will contain astrometric offsets for every measurement taken of every star, so-called `epoch astrometry' \citep{brown_highlight21}. This is expected to significantly improve prospects for exoplanet detection and characterisation \citep{sozzetti24,feng24}.\\
To test the prospects of this new data before DR4, we simulate a system with a single giant planet and produce expected astrometric offsets measured by \textit{Gaia}, fitted track parameters and RUWE.  These are then used to recover the simulated orbital parameters.  For this example, we assume a stellar mass of 1\,M$_{\odot}$, a planet mass of 10\,M$_{\mathrm{J}}$, a period of 3.5\,years, eccentricity of 0.3, $i$, $\Omega$ and $\omega$ of 65\,$^{\circ}$, 30\,$^{\circ}$ and 240\,$^{\circ}$ respectively.  Periastron time is set to 2016.0.  The true distance is 20\,pc and the proper motion is 5\,mas/yr in the R.A. direction and 84\,mas/yr in the Dec direction.  This example is designed to resemble $\beta$-Pic, having the same position, distance and proper motion, but with a smaller, fainter host star which makes this a more suitable target for study.  The expected observation times and scanning angles for DR4, spanning 66 months were obtained using the \textit{Gaia} Observing Scheduling Tool (GOST).  The simulated epoch data expected in DR4 are shown in Figure~\ref{fig:epoch_sim}.\\
The lower panel of Figure~\ref{fig:epoch_sim} shows that differences between the measured positions and fitted track are significantly larger than the assumed uncertainty of 0.1\,mas.\\
This simulated system was analysed, first using the RUWE and fitted track parameters assuming DR3 data and then using the DR4 epoch data shown in Figure~\ref{fig:epoch_sim}.  The corner plot for planet mass, period, eccentricity and inclination is shown in Figure~\ref{fig:bPic_epoch}, using DR3 and DR4 data, with the simulated values marked for comparison.
The result in Figure~\ref{fig:bPic_epoch} demonstrates significant improvement when the epoch data from DR4 are used.  The planet mass, which was previously overestimated, is now more tightly constrained and matches up closely with the simulated value.  The period and inclination are now closer to the simulated values.  Eccentricity could not be properly constrained without epoch data and favoured values near the upper cutoff of 0.8  With epoch data, the eccentricity is closer to the simulated value of 0.3.  Neither of these analyses used any imaging data (real or simulated) so this demonstrates a significant improvement when epoch data are used.  The resultant distributions of orbital angles are shown in Figure~\ref{fig:bPic_epoch_ang}.\\
The result shown in Figure~\ref{fig:bPic_epoch_ang} demonstrates all orbital angles are more tightly constrained when epoch astrometry is used.  The longitude of the ascending node is close to the simulated value of 30\,$^{\circ}$ when epoch data are used whereas the argument of periastron is still $\sim$35\,$^{\circ}$ away from the true value of 240$^{\circ}$.  However there is still a significant improvement from DR3.  A summary of these results is presented in Table~\ref{tab:bPic_summary}.
\begin{table}
    \centering
    \begin{tabular}{c|c|c|c|c|c|c}
        Parameter & Simulated Value & DR3 & DR4\\\hline
        Mass (M$_{\rm{J}}$) & 10.0 & 30.5$^{+0.9}_{-1.0}$ & 10.1$\pm$0.1\\\\
        Period (years) & 3.5 & 3.08$^{+0.16}_{-0.15}$ & 3.45$\pm$0.02\\\\
        Eccentricity & 0.3 & 0.80$^{+0.00}_{-0.01}$ & 0.37$\pm$0.03\\\\
        Inclination ($^{\circ}$) & 65 & 112$\pm$2 & 68$^{+1}_{-2}$\\\\
        $\Omega$ ($^{\circ}$) & 30 & 48$\pm$3 & 35$\pm$2\\\\
        $\omega$ ($^{\circ}$) & 240 & 338$\pm$4 & 205$\pm$2\\\\
    \end{tabular}
    \caption{Comparison of DR3 and DR4 results to simulated values for simulated single planet system, as shown in Figures~\ref{fig:bPic_epoch} and~\ref{fig:bPic_epoch_ang}.}
    \label{tab:bPic_summary}
\end{table}\\
\section{Expected Planet Yield From \textit{Gaia} Survey}
\begin{figure}
    \centering\includegraphics[width=1.0\linewidth]{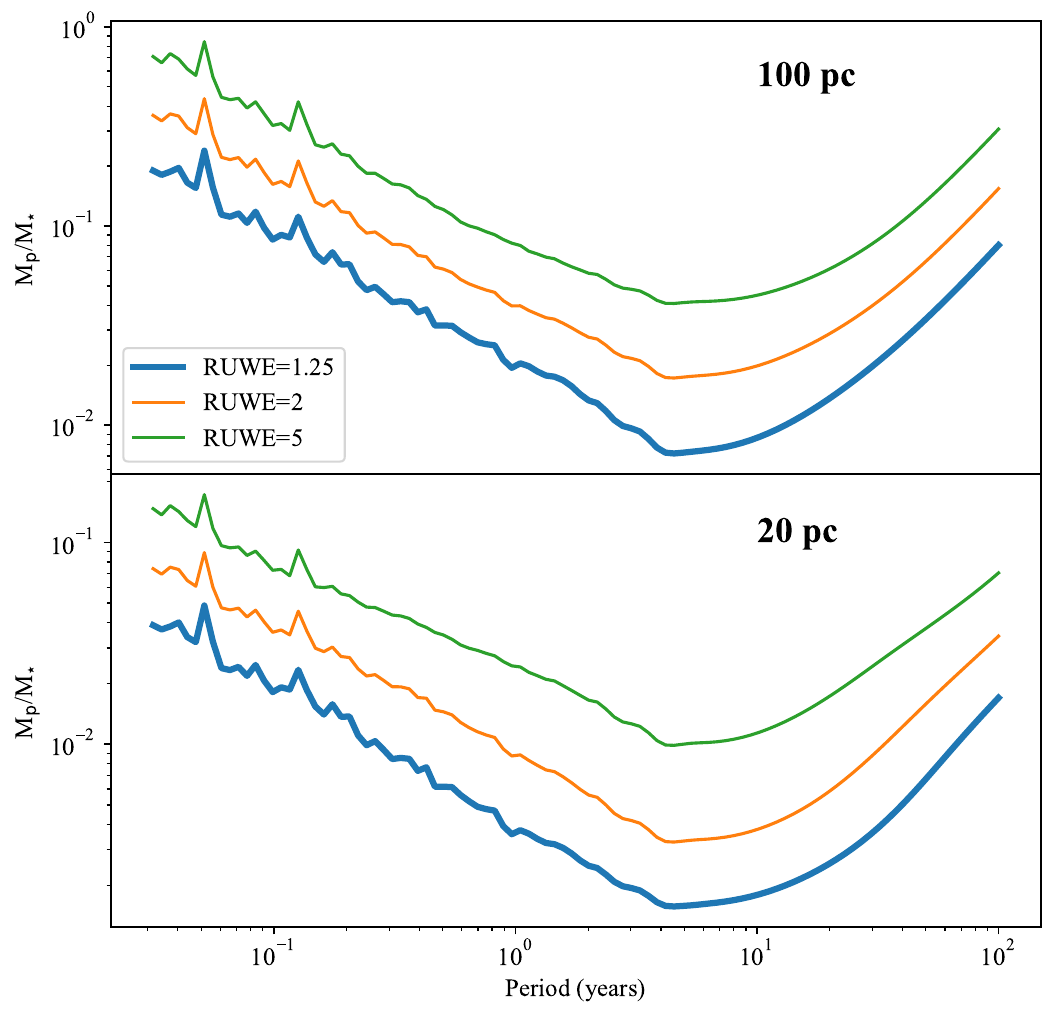}
    \caption{Minimum detectable mass ratio at a distance of 100\,pc (top) and 20\,pc (bottom) as a function of period assuming RUWE thresholds of 1.25, 2.0 and 5.0.}
    \label{fig:detectability}
\end{figure}

\begin{figure}
    \centering\includegraphics[width=1.0\linewidth]{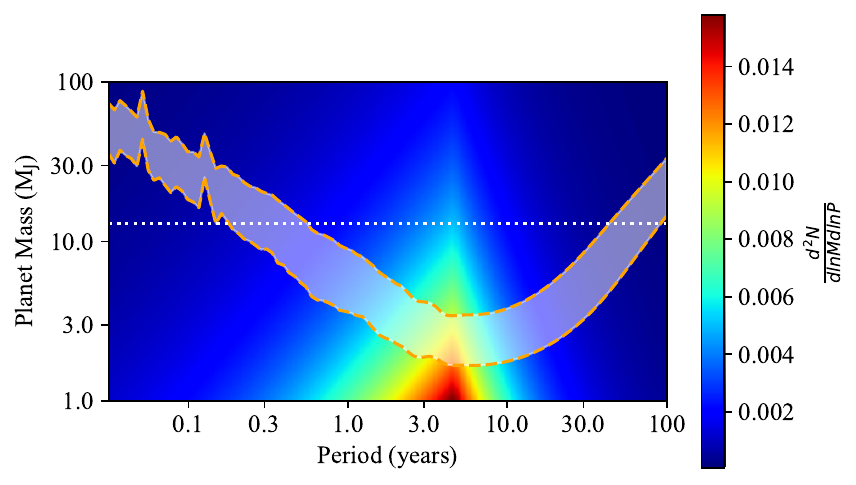}
    \caption{Distribution of planet mass and period from \citet{fernandes2019hints}, minimum detectable planet mass and maximum possible planet mass, assuming a stellar mass of 1\,M$_{\odot}$ at a distance of 20\,pc and measured RUWE of 2 (dashed lines).  Colour-scale represents derivative of total number of planets ($N$) with respect to mass and period in logspace.}  Shaded region covers range of possible detectable planets for this \textit{Gaia} source.  Dotted line is planet mass threshold of 13\,M$_{\mathrm{J}}$.
    \label{fig:distribution}
\end{figure}

\begin{figure}
    \centering\includegraphics[width=1.0\linewidth]{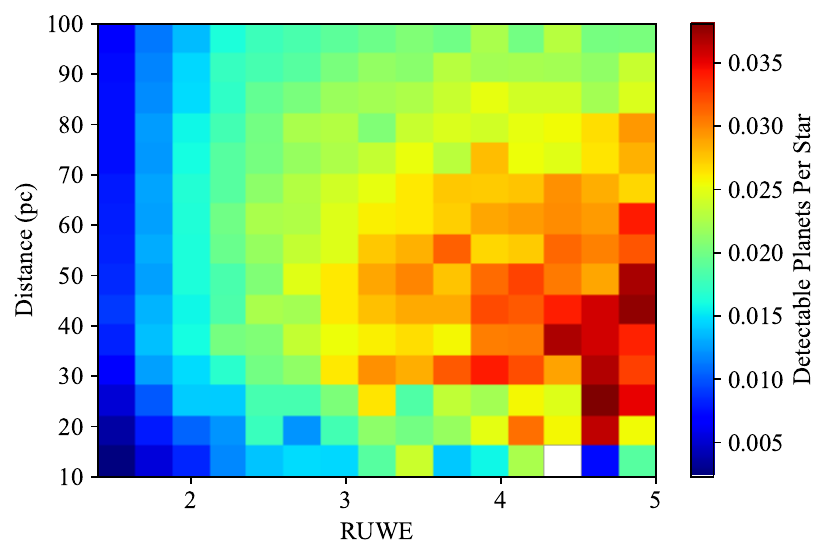}
    \caption{Average expected planet detections per star as a function of RUWE and distance.}
    \label{fig:detections}
\end{figure}

The previous sections have shown that a high RUWE can be indicative of a hidden companion but bright sources can make this method unreliable.  When planning future searches for planets, it is important to consider what types of planets are likely to be detected by \textit{Gaia}.\\
In order to investigate this, we simulated a set of systems at different masses and periods and calculated the average RUWE across multiple eccentricities, inclinations and other orbital parameters.  If we assume a planet is detectable if RUWE is above some `threshold' value, it is possible to determine the minimum planet mass detectable at a given period.  This minimum detectable mass would vary depending on stellar mass but this relation is approximately linear.  For this reason, we instead plot the minimum detectable mass ratio between planet and star as a function of period.  These minimum mass ratios are shown in Figure~\ref{fig:detectability} for RUWE thresholds of 1.25, 2.0 and 5.0 assuming distances of 100\,pc (top panel) and 20\,pc (bottom panel) and a time series from DR4.
As expected, the detectable mass ratio is minimised at periods of $\sim$5.5\,years as this is the length of \textit{Gaia} observations in DR4 and where we will be the most sensitive.  At these periods, even with a RUWE threshold of 2 we would expect to detect companions with a mass ratio as low as 0.004 which, for Sun-like stars, sits comfortably below the planet mass threshold of $\sim$13\,M$_{\mathrm{J}}$.\\
The RUWE threshold indicating a companion varies according to sky position and can be anywhere from 1.15 to 1.37 \citep{castro24}.  For the purpose of the example shown in Figure~\ref{fig:distribution}, we use 1.25 as a threshold RUWE, which is the sky-averaged threshold for detecting a companion with DR3 \citep{penoyre22}.  We use this value because the threshold for companion detection with DR4 is not yet known.  From this threshold, we calculate the minimum detectable planet mass as a function of period which is represented by the thick bottom curve in Figure~\ref{fig:detectability}.  For a given \textit{Gaia} source, we can also use the measured RUWE to estimate the maximum planet mass as a function of period.  These masses can then be compared to the expected giant planet distribution given by \citet{fernandes2019hints}.  This distribution takes the form of a broken power law:
\begin{equation}
    \frac{d^{2}N}{dlnMdlnP} \propto M^{\alpha}P^{\beta}.
\end{equation}
where $\alpha=0.45$ and the value of $\beta$ varies on either side of the snow line.  For this study, we use the asymmetric distribution where $\beta=0.53$ for periods less than 2,075\,days and $\beta=-1.22$ for periods greater than 1717\,days.\\
An example of the comparison between detectable mass and this planet distribution is shown in Figure~\ref{fig:distribution} assuming a \textit{Gaia} source with a mass of 1\,M$_{\odot}$ at a distance of 20\,pc and a RUWE of 2.
Figure~\ref{fig:distribution} shows a clear region in period-space in which planets orbiting a 1\,M$_{\odot}$ star at 20\,pc can be detected.  Between periods of $\sim$0.3\,years and 20\,years the minimum detectable mass is significantly below the 13\,M$_{\mathrm{J}}$ threshold and, with a RUWE of 2, the maximum mass is also below this threshold.  Integrating the expected distribution from \citet{fernandes2019hints} shows that 5\,\% of stars should host planets $>$1\,M$_{\mathrm{J}}$ in this period range.\\
In order to investigate the expected planet yield from \textit{Gaia}, a sample of sources was selected with RUWE between 1.4 and 5.  The reason for the upper limit of 5 was that beyond this, the high RUWE can usually be explained by a secondary star or a brown dwarf and is unlikely to be solely caused by a planet.  The sample also restricted to sources not featured in the non-single star (NSS) table with a distance range from 10--100\,pc.  The resultant sample size was 27,310 sources.  For each source, the stellar mass was estimated from the absolute G magnitude and the minimum detectable mass was calculated as a function of period assuming a threshold RUWE of 1.4, as well as the maximum mass using the measured RUWE, similar to the dashed lines in Figure~\ref{fig:distribution}.  The planet distribution was integrated over all periods and a mass range from the minimum mass to the maximum mass or 13\,M$_{\mathrm{J}}$ (whichever was lower) to estimate the number of detectable planets for that source.  The average planet yield as a function of RUWE and distance is shown in Figure~\ref{fig:detections}.
The expected yield as a function of RUWE and distance demonstrates an increase in detections with higher RUWE at smaller distances.  The dependence on distance is due to the fact that, at larger distances, a larger companion mass is required to explain a high RUWE which is more likely to be higher than the planet mass threshold.  Averaging over all distances, we find that approximately 2\,\% of sources with RUWE greater than 3.5 are expected to host a detectable planet whereas only 1\,\% of sources with RUWE$\sim$2 are expected to host a detectable planet.  In total, 311 of these sources are expected to host a detectable planet.  For the other sources, the RUWE can be explained by multiple companions or brown dwarfs on wider orbits.  Section~\ref{sec:epoch} shows that we will require time-series data in order to properly constrain the properties of any potential planets.
\section{Summary and Conclusions}
In this paper, we have shown how the measured RUWE of a source can be indicative of a hidden companion and how this value depends strongly on the mass and period of the companion.  We have demonstrated a Bayesian analysis technique for constraining the properties of a potential companion.  However, with \textit{Gaia} DR3 data, this method is most reliable if additional radial velocity or imaging data are used.  For this reason, the method can currently only be accurately applied to sources with known planets.\\
By combining this astrometric method with radial velocity measurements, we are able to constrain a planet's mass and inclination and have identified two known companions, previously categorised as planets, with upper mass limits in the brown dwarf range.  Although this method has not been applied to all systems, the RUWE values of several targets suggests companions with masses too high to be classified as planets, though this could also be explained by calibration errors.\\
This analysis method was attempted with the directly imaged planet $\beta$-Pictoris c.  Our results suggest that the star's RUWE can be explained by the two known 10\,M$_{\rm{J}}$ planets but we acknowledge there may be other explanations, due to the star's brightness.  We are, however, optimistic that more planets in more suitable systems will be directly imaged in the future. By combining \textit{Gaia}'s limits of planet detection with expected planet distributions, we have quantified the average probability that a star with a given RUWE and distance hosts a detectable planet.  However, we stress that high RUWE can also be explained by calibration errors and are unable to draw a definite conclusion.\\
Finally, by simulating a nearby planet around a Sun-like star, we show how access to epoch astrometry in \textit{Gaia} DR4 will allow us to constrain orbital parameters of companions without the need for other data such as radial velocity and imaging.  This will revolutionise the hunt for exoplanets by astrometry, allowing us to probe an area of period-space difficult to access with other methods.  With enough epoch data, it will also be possible to constrain the properties of multiple planets in the same system.  It is also expected that the extra information contained in epoch astrometry will also make it easier to distinguish between a companion and a calibration error.  We already have access to the RUWE which can help us identify targets of interest for future studies and with these analysis tools, with the release of DR4, we will be able to apply this method to identify and constrain parameters of potential planets.  With advances in direct imaging and interferometry, it may also be possible to perform followup observations on young systems in order to understand more about giant planet formation.
\section*{Data Availability}
The data underlying this article are available from the corresponding author on reasonable request.  The \texttt{pystan} program used to constrain companion parameters is available at \href{https://github.com/awallace142857/gaia_astro_fit}{https://github.com/awallace142857/gaia\_astro\_fit}.
\section*{Acknowledgements}
This work was funded by the Australian Research Council Discovery Grant DP210100018.\\
This work presents results from the European Space Agency (ESA) space mission Gaia. Gaia data are being processed by the Gaia Data Processing and Analysis Consortium (DPAC). Funding for the DPAC is provided by national institutions, in particular the institutions participating in the Gaia MultiLateral Agreement (MLA). The Gaia mission website is \href{https://www.cosmos.esa.int/gaia}{https://www.cosmos.esa.int/gaia}. The Gaia archive website is \href{https://archives.esac.esa.int/gaia}{https://archives.esac.esa.int/gaia}.




\renewcommand\refname{References}
\bibliographystyle{mnras}
\bibliography{references} 








\bsp	
\label{lastpage}
\end{document}